\documentclass[prc,amsmath,onecolumn,floatfix,superscriptaddress,showpacs,showkeys,nofootinbib,preprint]{revtex4}
\usepackage{epsfig}
\newcommand{\eq}[1]{\begin{align} #1 \end{align}}

\begin{document}

\title{Particle Number Fluctuations and Correlations \\in Nucleus-Nucleus Collisions
}\texttt{}

\author{V.P.~Konchakovski}
\affiliation{Bogolyubov Institute for Theoretical Physics, Kiev, Ukraine}
\affiliation{Helmholtz Research School, University of Frankfurt, Frankfurt, Germany}
\author{M.~Hauer}
\affiliation{Helmholtz Research School, University of Frankfurt, Frankfurt, Germany}
\author{M.I.~Gorenstein}
\affiliation{Frankfurt Institute for Advanced Studies, Frankfurt, Germany}
\affiliation{Bogolyubov Institute for Theoretical Physics, Kiev, Ukraine}
\author{E.L.~Bratkovskaya}
\affiliation{Institute f\"ur Theoretische Physik, Frankfurt, Germany}

\begin{abstract}

Particle number fluctuations and correlations in
nucleus-nucleus collisions at SPS and RHIC energies are
studied within the statistical hadron-resonance gas model in
different statistical ensembles and in the Hadron-String-Dynamics
(HSD) transport approach. Event-by-event fluctuations of the
proton to pion and kaon to proton number ratios are calculated in
the HSD model for the samples of most central collision events and
compared with the available experimental data. The role of the
experimental acceptance and centrality selection is  discussed.
\end{abstract}


\maketitle


\section{Introduction}

The study of event-by-event fluctuations in high energy
nucleus-nucleus  collisions opens new possibilities to investigate
the phase transition between hadronic and partonic matter as well
as the QCD critical point (see the reviews \cite{fluc1}). In our
recent paper \cite{KPi} we have presented a systematic study of the
particle number fluctuations  for pions $\pi=\pi^++\pi^-$ and
kaons $K=K^++K^-$ and their correlations in the statistical model
(SM) in different ensembles and in the Hadron-String-Dynamics
(HSD) transport model~\cite{HSD}. In the present paper we continue this study
and consider the fluctuations of the number of protons
p$=p+\bar{p}$, the $\pi$-$p$ and $K$-$p$ correlations, and
fluctuations of the $p/\pi$ and $K/p$ particle number ratios. The
HSD results for the fluctuations in the particle number ratios are
compared with NA49 data for Pb+Pb collisions at SPS energies
\cite{NA49ratio} for the proton to pion ratio as well as preliminary STAR data
for Au+Au collisions at RHIC energies  for the proton to pion
\cite{STAR-PPi} and kaon to proton  \cite{KP} ratios.

The paper is organized as follows. In Section II the observables for
particle number fluctuations and correlations are introduced. In
Section III the  SM and HSD  results for the scaled variances and
correlation coefficients of particle number fluctuations are
presented for central nucleus-nucleus collisions at SPS and
RHIC energies. 
In Section IV the HSD transport model results for the fluctuations of the $p/\pi$ and $K/p$ ratios 
are presented and compared to the available data on proton to pion ratio fluctuations. 
The role of the centrality selection and experimental acceptance is discussed using  the HSD results.
A summary closes the paper in Section V.

\section{Observables for Particle Ratio Fluctuations}

We define the covariance for particle species $A$ and $B$ as:
\eq{ \label{cov}
\Delta \left(N_A,N_B\right)~\equiv~\langle \Delta N_A
\Delta N_B\rangle~=~\langle N_A N_B\rangle
~-~\langle N_A\rangle \langle N_B\rangle~,
}
where $\langle N_A\rangle$ is the event-by-event average of the particle multiplicity
$N_A$ and $\Delta N_A\equiv N_A-\langle N_A\rangle$. The scaled variance $\omega_A$
and correlation coefficient $\rho_{AB}$ are defined by:
\eq{\label{omega}
\omega_A ~\equiv~\frac{\Delta\left(N_A,N_A\right)}{\langle
N_A\rangle}~,~~~~
\rho_{AB}~\equiv~\frac{\langle\Delta N_A~\Delta N_B\rangle}{\left[\langle\left(\Delta
N_A\right)^2\rangle~\langle\left(\Delta N_B\right)^2\rangle\right]^{1/2}}~.
}
Furthermore, the fluctuations in the ratio $R_{AB}\equiv N_A/N_B$
will be characterised by:
\eq{\label{sigma-def} \sigma_{AB}^2~\equiv~\frac{\langle
\left(\Delta R_{AB}\right)^2\rangle} {\langle R_{AB}\rangle^2 } ~.
}
One finds to second order in $\Delta N_A/\langle N_A \rangle$
and $\Delta N_B/\langle N_B \rangle$:
\eq{
\label{sigma}
\sigma_{AB}^2~\cong~
 ~\frac{\omega_A}{\langle N_A\rangle}~+~\frac{\omega_B }{\langle
N_B\rangle}~-~2\rho_{AB}~\left[\frac{\omega_A \omega_B}{\langle
N_A\rangle\langle N_B\rangle}\right]^{1/2}~.
}

We recall that the experimental data for $N_A/N_B$ fluctuations are usually presented
in terms of the so called dynamical fluctuations 
\cite{VKR}\footnote{Other dynamical measures, such as $\Phi$ \cite{GM,M} and $F$ \cite{Koch}, may be also used.}
\eq{\label{sigmadyn}
\sigma_{AB}^{dyn}~\equiv~\texttt{sign}\left(\sigma_{AB}^2~-
~\sigma^2_{AB,mix}\right)\left|\sigma_{AB}^2~-~\sigma^2_{AB,mix}\right|^{1/2}~,
}
where $\sigma_{AB}^2$ is defined by Eq.~(\ref{sigma}) while
$\sigma^2_{AB,mix}$ corresponds to the {\it mixed events} background (see Ref.~\cite{KPi}):
\eq{  
\sigma^2_{AB,mix}~=~ \frac{1}{\langle N_A \rangle }~+~\frac{1}{\langle N_B \rangle}~. 
\label{Dmix}
}
%

\section{Particle Number Fluctuations and Correlations}

\begin{figure}[ht!]
\epsfig{file=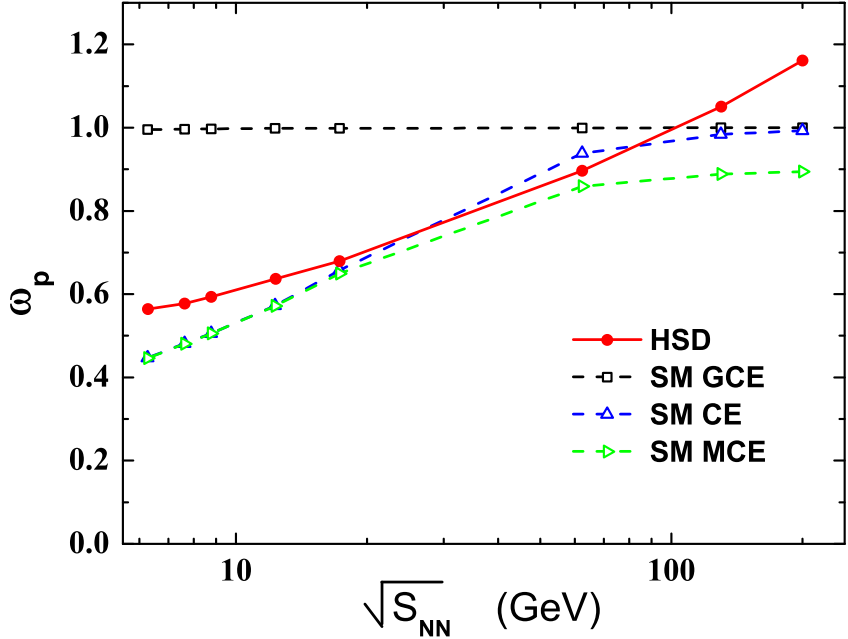,width=12cm}
\caption{(Color online) 
The SM results in the GCE, CE, and MCE ensembles and the HSD
results (impact parameter $b=0$) are presented for the scaled variance
$\omega_{p}$ in Pb+Pb (Au+Au) collisions at different
c.m. energies $\sqrt{s_{NN}}$.}
\label{fig1}
\end{figure}

\begin{figure}[ht!]
\epsfig{file=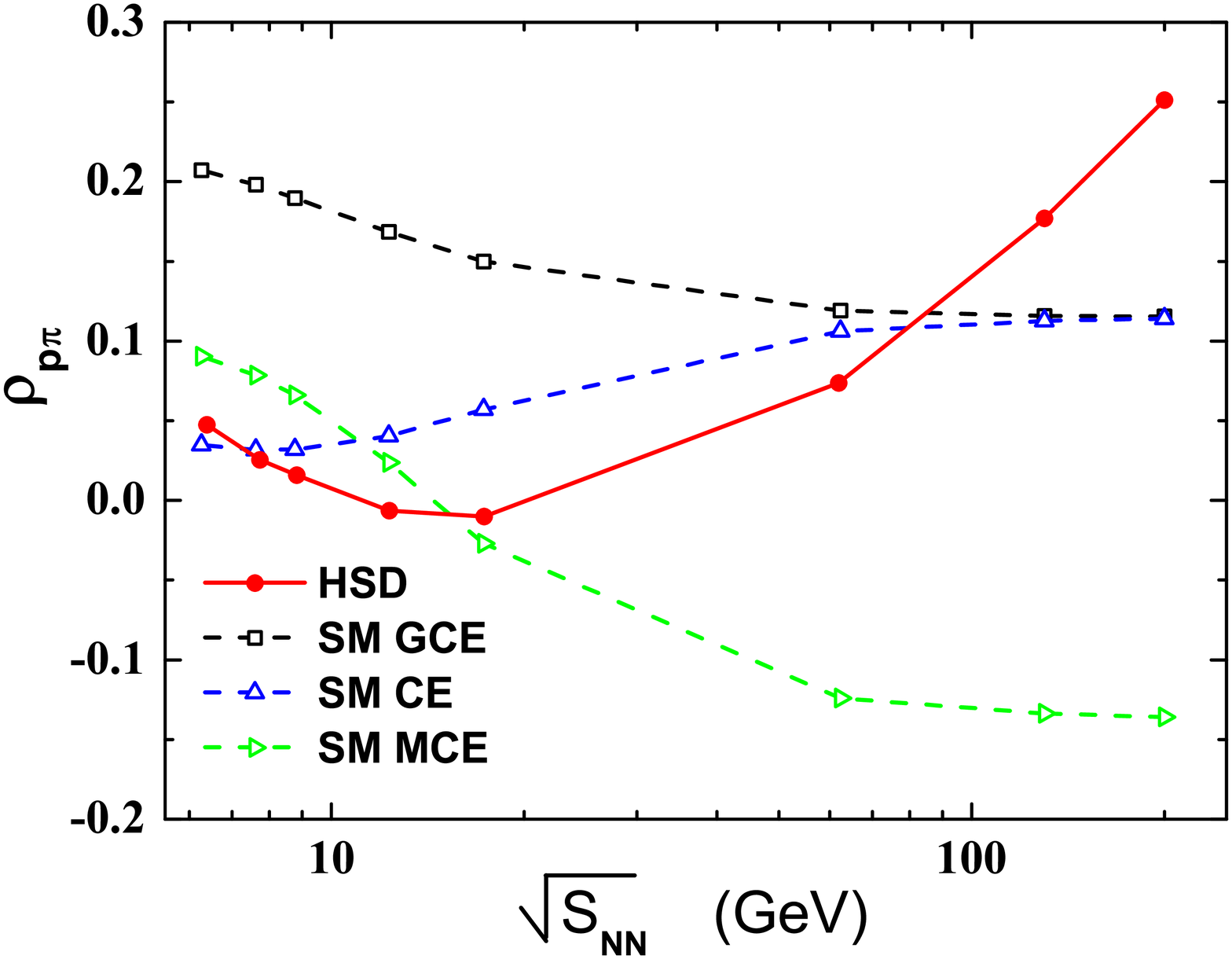,width=12cm}
\epsfig{file=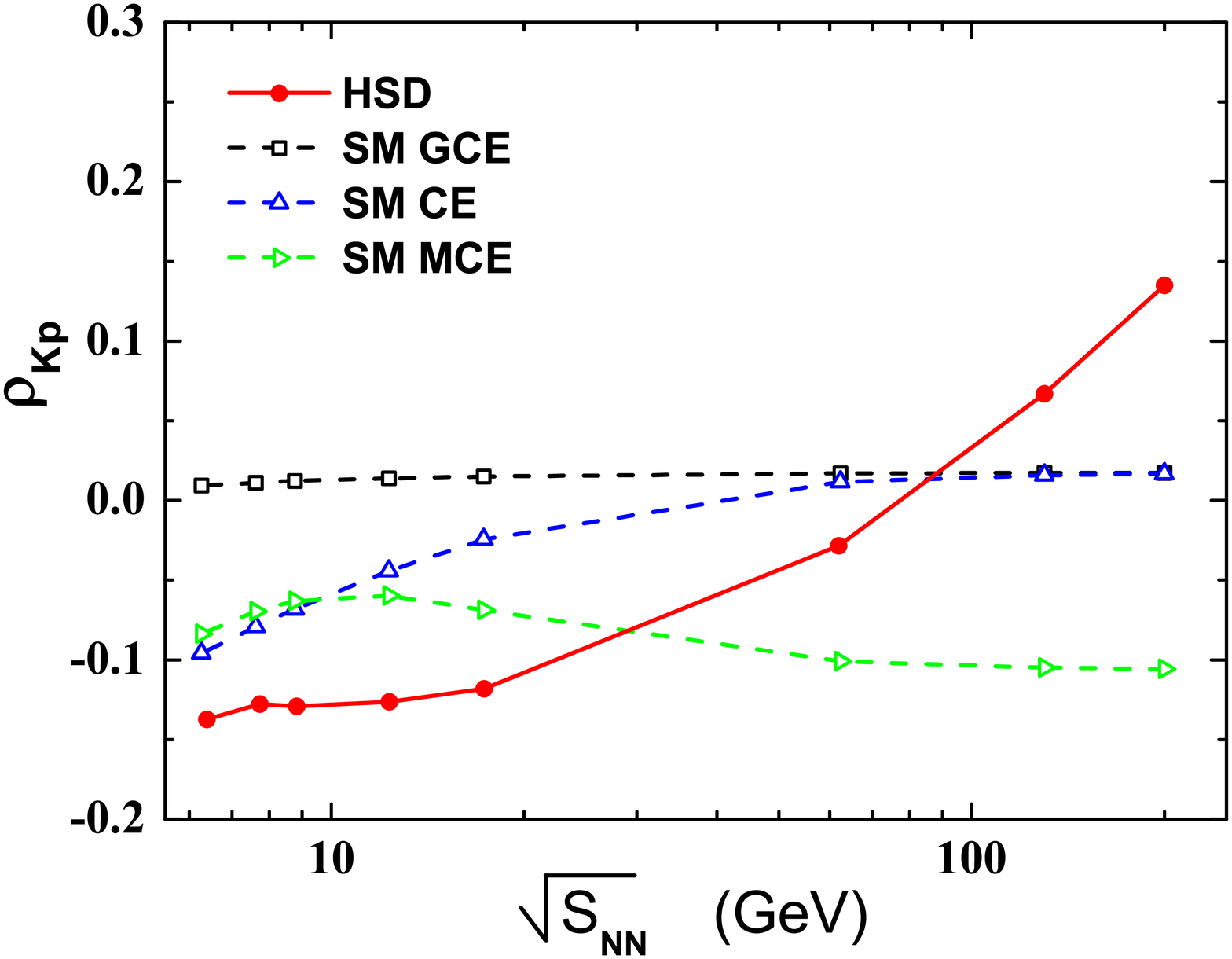,width=12cm}
\caption{(Color online) 
The SM results in the GCE, CE, and MCE ensembles and the HSD
results (impact parameter $b=0$) are presented for the correlation
parameters $\rho_{p\pi}$ ({\it upper pannel}) and $\rho_{Kp}$
({\it lower panel}) in Pb+Pb (Au+Au) collisions at different c.m.
energies $\sqrt{s_{NN}}$.}
\label{fig2}
\end{figure}

In this section we present the results of the SM and HSD for
the particle number fluctuations and correlations in central nucleus-nucleus collisions. 
The procedure of the calculations is essentially the same as in our previous paper \cite{KPi}.
The SM and HSD results for scaled variances of kaons $K=K^++K^-$, pions
$\pi=\pi^++\pi^-$, and protons p$=p+\bar{p}$, and the correlation coefficients
$\rho_{K\pi}$, $\rho_{\pi p}$, and $\rho_{Kp}$ (defined by Eq.~(\ref{omega})) are presented in Tables I and II.
Figures~\ref{fig1} and \ref{fig2} show $\omega_p$ and $\rho_{p\pi}$, $\rho_{Kp}$, respectively. 
We note the results for $\omega_K$, $\omega_{\pi}$ and $\rho_{K\pi}$ have been presented in Ref.~\cite{KPi}.

\begin{table}[ht!]
 \begin{center}
 {\footnotesize
 \begin{tabular}{||c||c|c||c||c|c|c||c|c|c||c|c|c||c|c|c||}\hline
 \ $\sqrt{s_{NN}}$ &\ $T$ &\ $\mu_B$ &  $n_p$\ & \multicolumn{3}{c||}{GCE} & \multicolumn{3}{c||}{CE} & \multicolumn{3}{c||} {MCE} \\\hline
 \ [ GeV ] &\ [ MeV ] &\ [ MeV ] &\  $[ fm^{-3} ]$&\ $\omega_p$ &\ $\rho_{p\pi}$ &\ $\rho_{Kp}$ &\ $\omega_p$ &\ $\rho_{p\pi}$ &\ $\rho_{Kp}$ &\ $\omega_p$ &\ $\rho_{p\pi}$ &\ $\rho_{Kp}$\\\hline\hline
 \ 6.27&\ 130.7&\ 482.4&\ 0.048&\ 0.995&\ 0.207&\ 0.009&\ 0.447&\ 0.035&\ -0.096&\ 0.446&\  0.090&\ -0.084\\
 \ 7.62&\ 138.3&\ 424.6&\ 0.047&\ 0.996&\ 0.198&\ 0.011&\ 0.481&\ 0.032&\ -0.079&\ 0.480&\  0.078&\ -0.070\\
 \ 8.77&\ 142.9&\ 385.4&\ 0.046&\ 0.997&\ 0.190&\ 0.012&\ 0.506&\ 0.032&\ -0.068&\ 0.505&\  0.066&\ -0.063\\
 \ 12.3&\ 151.5&\ 300.1&\ 0.041&\ 0.998&\ 0.168&\ 0.014&\ 0.573&\ 0.040&\ -0.044&\ 0.572&\  0.024&\ -0.060\\
 \ 17.3&\ 157  &\ 228.6&\ 0.037&\ 0.999&\ 0.150&\ 0.015&\ 0.658&\ 0.057&\ -0.024&\ 0.649&\ -0.027&\ -0.069\\
 \ 62.4&\ 163.1&\ 72.7 &\ 0.027&\ 1.000&\ 0.119&\ 0.017&\ 0.938&\ 0.106&\  0.012&\ 0.859&\ -0.124&\ -0.101\\
 \ 130 &\ 163.6&\ 36.1 &\ 0.026&\ 1.000&\ 0.116&\ 0.017&\ 0.984&\ 0.113&\  0.016&\ 0.888&\ -0.134&\ -0.105\\
 \ 200 &\ 163.7&\ 23.4 &\ 0.026&\ 1.000&\ 0.115&\ 0.017&\ 0.993&\ 0.114&\  0.017&\ 0.894&\ -0.136&\ -0.106\\
     \hline
 \end{tabular}
 }
 \caption{The chemical freeze-out parameters $T$ and $\mu_B$ for central Pb+Pb (Au+Au) collisions
along the chemical freeze-out line~\cite{fo-line}.
The hadron-resonance gas model results are presented for
the number density of $p+\bar{p}$,
(particle number densities are the same in all statistical ensembles in the large volume limit),
scaled variance $\omega_{p}$, and correlation parameter $\rho_{p\pi}$ and $\rho_{Kp}$
in the GCE, CE, and MCE. The SM results for $n_\pi$, $n_K$,
 $\omega_{\pi}$, $\omega_K$ and $\rho_{K\pi}$
are presented in Table I of Ref.~\cite{KPi}
}
 \end{center}
\end{table}
%

%
\begin{table}[ht!]
 \begin{center}
 {\footnotesize
 \begin{tabular}{||c||c|c|c|c|c|c|c|c|c||}\hline
 \ $\sqrt{s_{NN}}$ & \multicolumn{9}{c||}{HSD full acceptance} \\\hline
 \ [ GeV ] &\ $\langle N_\pi\rangle $ &\ $\langle N_K\rangle$ &\ $\langle N_p\rangle$
  &\ $\omega_{\pi}$ &\ $\omega_K$ &\ $\omega_p$ &\ $\rho_{K\pi}$ &\ $\rho_{p\pi}$  &\ $\rho_{Kp}$ \\\hline\hline
     \ 6.27&\ 612.03&\ 43.329 &\ 181.83 &\ 0.961&\ 1.107&\ 0.506 &\  -0.091&\ 0.048  &\ -0.137 \\
     \ 7.62&\ 732.11&\ 60.801 &\ 180.03 &\ 1.077&\ 1.141&\ 0.526 &\  -0.063&\ 0.025  &\ -0.128 \\
     \ 8.77&\ 823.71&\ 75.133 &\ 179.43 &\ 1.159&\ 1.168&\ 0.546 &\  -0.033&\ 0.016  &\ -0.129 \\
     \ 12.3&\ 1072.3&\ 116.44 &\ 180.81 &\ 1.378&\ 1.250&\ 0.596 &\  0.046&\  -0.006 &\ -0.126  \\
     \ 17.3&\ 1364.6&\ 165.52 &\ 186.97 &\ 1.619&\ 1.348&\ 0.641 &\  0.126&\  -0.010 &\ -0.118  \\
     \ 62.4&\ 2933.9&\ 449.29 &\ 240.53 &\ 3.006&\ 1.891&\ 0.863 &\   0.412&\ 0.074  &\ -0.029  \\
     \ 130 &\ 4304.2&\ 692.59 &\ 307.31 &\ 4.538&\ 2.378&\ 1.020 &\  0.557&\  0.177  &\ 0.067 \\
     \ 200 &\ 5204.0&\ 861.77 &\ 352.91 &\ 5.838&\ 2.765&\ 1.122 &\   0.634&\ 0.251  &\ 0.135 \\
 \hline
 \end{tabular}
 }
 \caption{The HSD results for the average multiplicities $\langle N_{\pi}\rangle$,
 $\langle N_K\rangle$, $\langle N_p\rangle$ and values of $\omega_{\pi}$, $\omega_K$, and
$\rho_{K\pi}$ for central (impact parameter $b=0$) Pb+Pb (Au+Au)
collisions at different c.m. energies $\sqrt{s_{NN}}$.}
 \label{Tab1_N}
 \end{center}
\end{table}


Let us first comment on the SM results for the particle number
fluctuations and correlations in different ensembles. The global
charge conservation laws suppress the particle number fluctuations
in the canonical ensemble (CE) in a comparison with the grand canonical ensemble (GCE). 
Exact energy conservation in the micro canonical ensemble (MCE) makes this suppression even stronger. 
This leads to the result that the MCE scaled variances $\omega_A$ become smaller than unity.
The contributions to the correlation parameter $\rho_{AB}$ in the
SM stem from two sources: resonance decays and global conservation
laws. Resonances decaying into pairs of particles of species $A$ and $B$ produce a
corresponding correlation and yield a positive contribution to $\rho_{AB}$. 
On the other hand, the conservation laws lead to an anticorrelation 
and thus to negative contributions to $\rho_{AB}$.

As seen from Figs.~\ref{fig1} and \ref{fig2} the HSD results for
$\omega_{p}$, $\rho_{p\pi}$ and $\rho_{Kp}$ (denoted by the solid lines)
are close to the CE and MCE results for low SPS energies. 
One may conclude  that  the influence of conservation laws is more
stringent at low collision energies. 
The same conclusion follows from the HSD results for $\omega_{\pi}$, $\omega_K$ and
$\rho_{K\pi}$ (see Figs.~1 and 2 in Ref.~\cite{KPi}). 
The HSD values for $\omega_A$ and $\rho_{AB}$ increase, however, at high collision energies
and a sizeable deviation of the HSD results from those in the MCE SM 
is observed with increasing energies for $\sqrt{s_{NN}}>200$~GeV.

\begin{figure}[ht!]
\epsfig{file=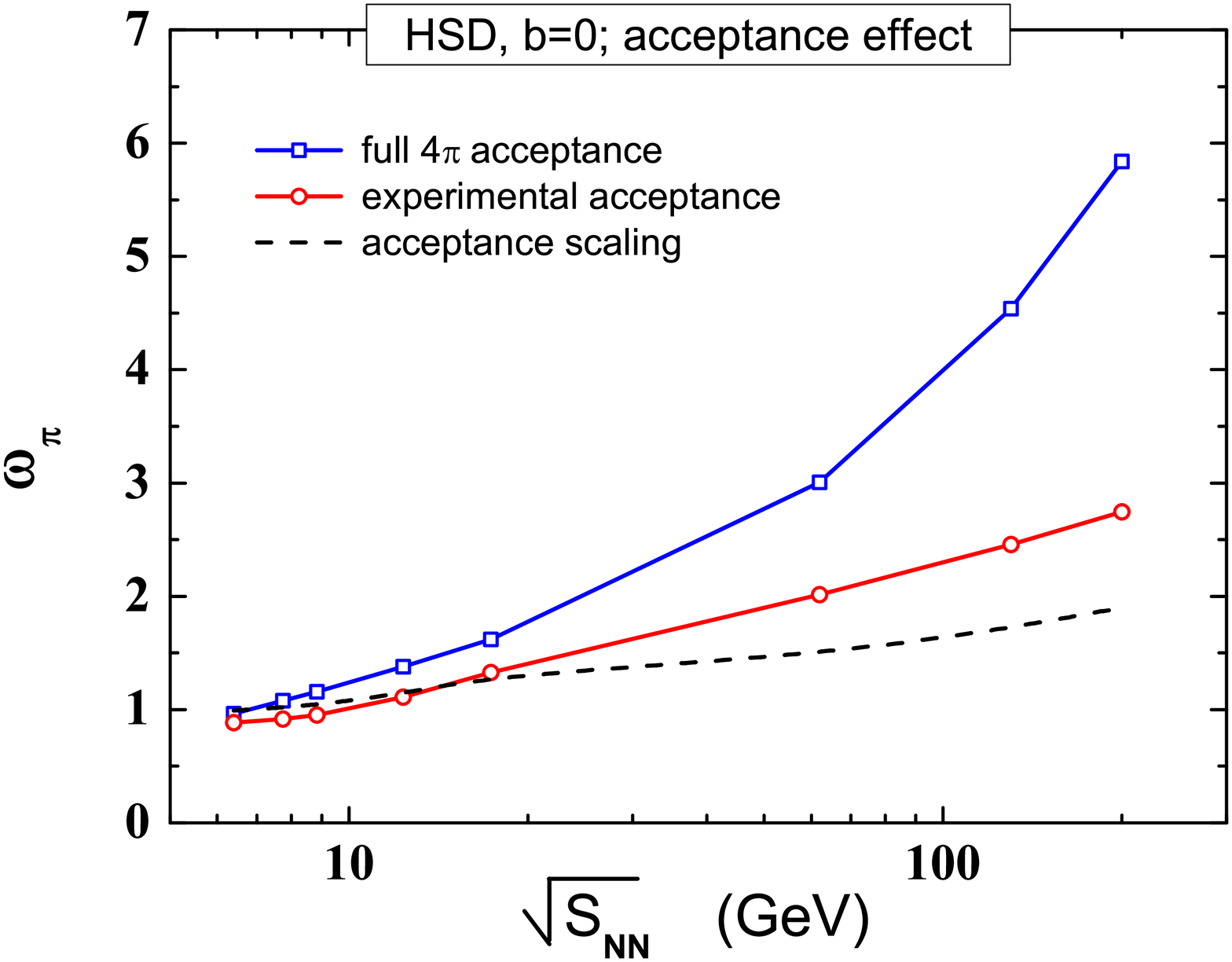,width=12cm}
\caption{(Color online)  
The HSD results for $\omega_\pi$  are presented  for Pb+Pb (Au+Au) collisions
with zero impact parameter ($b=0$) at different c.m. energies $\sqrt{s_{NN}}$. 
The upper solid line corresponds to the full $4\pi$-acceptance and the middle one to the experimental acceptance. 
The lower dashed line corresponds to the acceptance scaling Eq.~(\ref{acc}).}
\label{fig3}
\end{figure}

We point out again that important aspects of the event-by-event fluctuations in
nucleus-nucleus collisions are the dependence on the centrality
selection and experimental acceptance. 
We accordingly discuss the role of these effects using HSD results for the scaled variance
$\omega_{\pi}$ of the pion number fluctuations. 
In Fig.~\ref{fig3} the scaled variance $\omega_\pi$ (calculated within HSD) for the
full acceptance and for the experimental acceptance 
are shown in nucleus-nucleus collisions for zero impact parameter $b=0$ 
(see the next Section for details of the experimental acceptance). 
Introducing the probability $q$ of pion experimental
acceptance as the ratio of an average accepted to the total
multiplicities, $q=\langle N^{acc}_\pi\rangle/\langle
N^{tot}_\pi\rangle$, one finds:
\eq{\label{acc}
\omega_\pi^{acc}~=~1~-~q~+~q~ \omega_\pi^{full}~.
}
Equation~(\ref{acc}) connects the scaled variance
$\omega_\pi^{full}$ in the full $4\pi$ space with
$\omega_\pi^{acc}$ defined for the experimental acceptance. 
The acceptance scaling (\ref{acc}) assumes (see, e.g. Ref.~\cite{CE})
an absence of particle correlations in momentum space. 
Figure \ref{fig3} demonstrates that the acceptance scaling (\ref{acc}) 
underestimates the scaled variance $\omega_\pi^{acc}$ at RHIC energies.

The samples of collision events selected experimentally are 3.5\% of most central  collision
events in Pb+Pb collisions at the SPS energies and 5\% in Au+Au collisions at RHIC energies.
\begin{figure}[ht!]
\epsfig{file=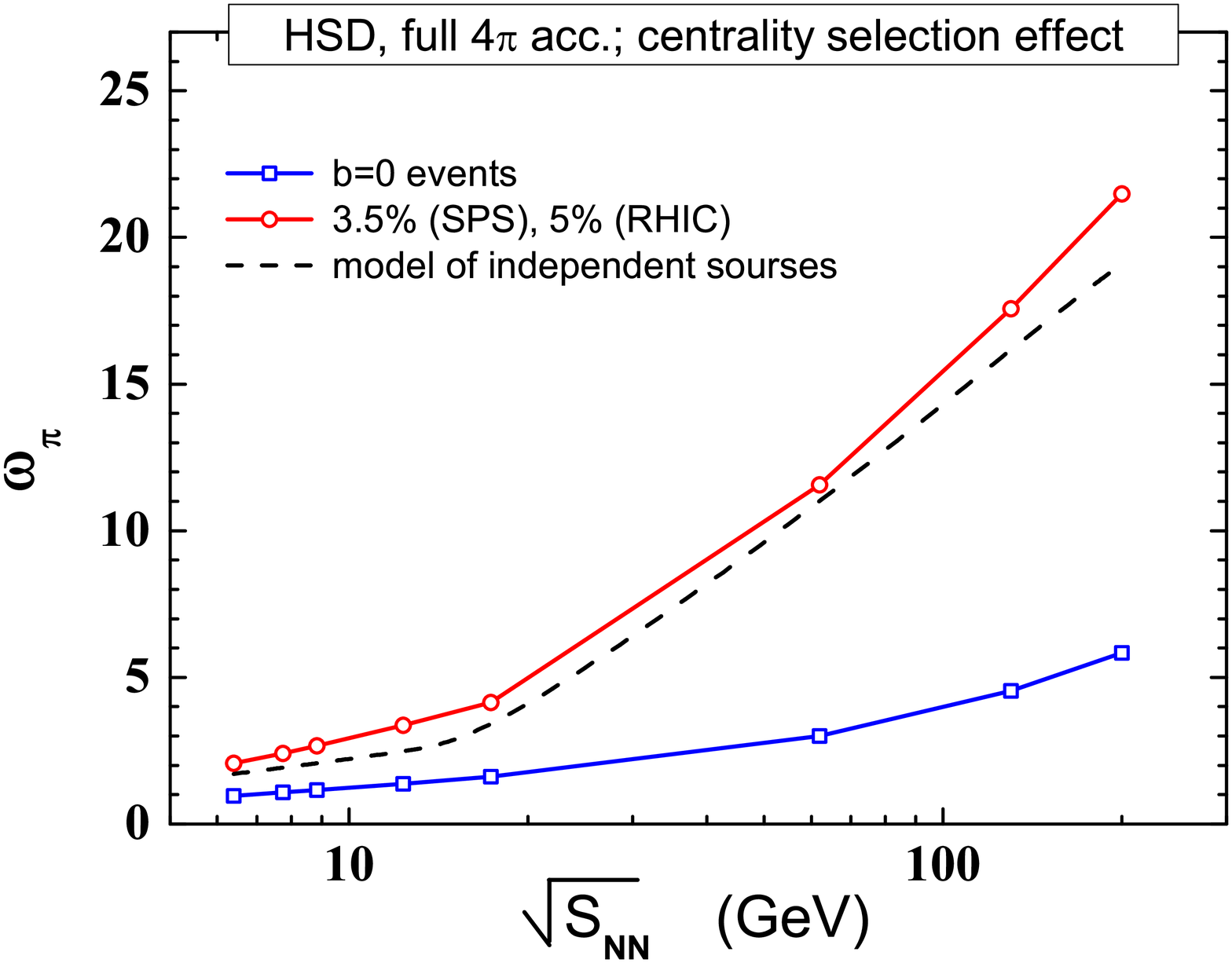,width=12cm}
\caption{(Color online)
The HSD results for $\omega_\pi$ for Pb+Pb (Au+Au) collisions
at different c.m. energies $\sqrt{s_{NN}}$ within the full $4\pi$-acceptance. 
The lower solid line corresponds to zero impact parameter ($b=0$) and 
the upper one to the experimentally selected samples of collision events. 
The dashed line reflects the model of independent sources (\ref{mis}).}
\label{fig4}
\end{figure}
Figure \ref{fig4} presents the HSD results for $\omega_\pi$ in
these samples of the most central events and their comparison with the
HSD results at zero impact parameter. 
One finds much larger values of $\omega_\pi$ in the centrality selected samples than for $b=0$. 
The effect is especially strong at RHIC energies.
This can be qualitatively understood within the model of
independent sources:
\eq{\label{mis}
\omega_{\pi}~=~\omega_{\pi}^s~+~n_{\pi}~ \omega_{P}~,
}
where $\omega_{\pi}^s$ is the scaled variance for pions produced
by one source, $\omega_P$ is the scaled variance for the
fluctuations of nucleon participants, and $n_{\pi}$ is the average
pion multiplicity per participating nucleon 
which increases monotonously with collision energy.
Collisions with zero impact parameter correspond to $\omega_P\cong 0$.
Thus, $\omega_{\pi}^s$ can be approximately taken as
$\omega_{\pi}$ at $b=0$. 
The HSD results correspond approximately
to $\omega_P\cong 0.5$ for the 3.5\% most central Pb+Pb collisions
at SPS energies and $\omega_P\cong 1$ for the 5\% most central
Au+Au collisions at RHIC energies. Please note that we used the
restrictions on impact parameter $b$ in the HSD calculations to
form the samples of most central events. 
The results of the model of independent sources (\ref{mis}) for $\omega_\pi$ are  shown by the
dashed line in Fig.~\ref{fig4} and are close to the actual values of
the HSD simulations for $\omega_\pi$.

\section{Fluctuations of Particle Number Ratios}

\begin{figure}[ht!]
\epsfig{file=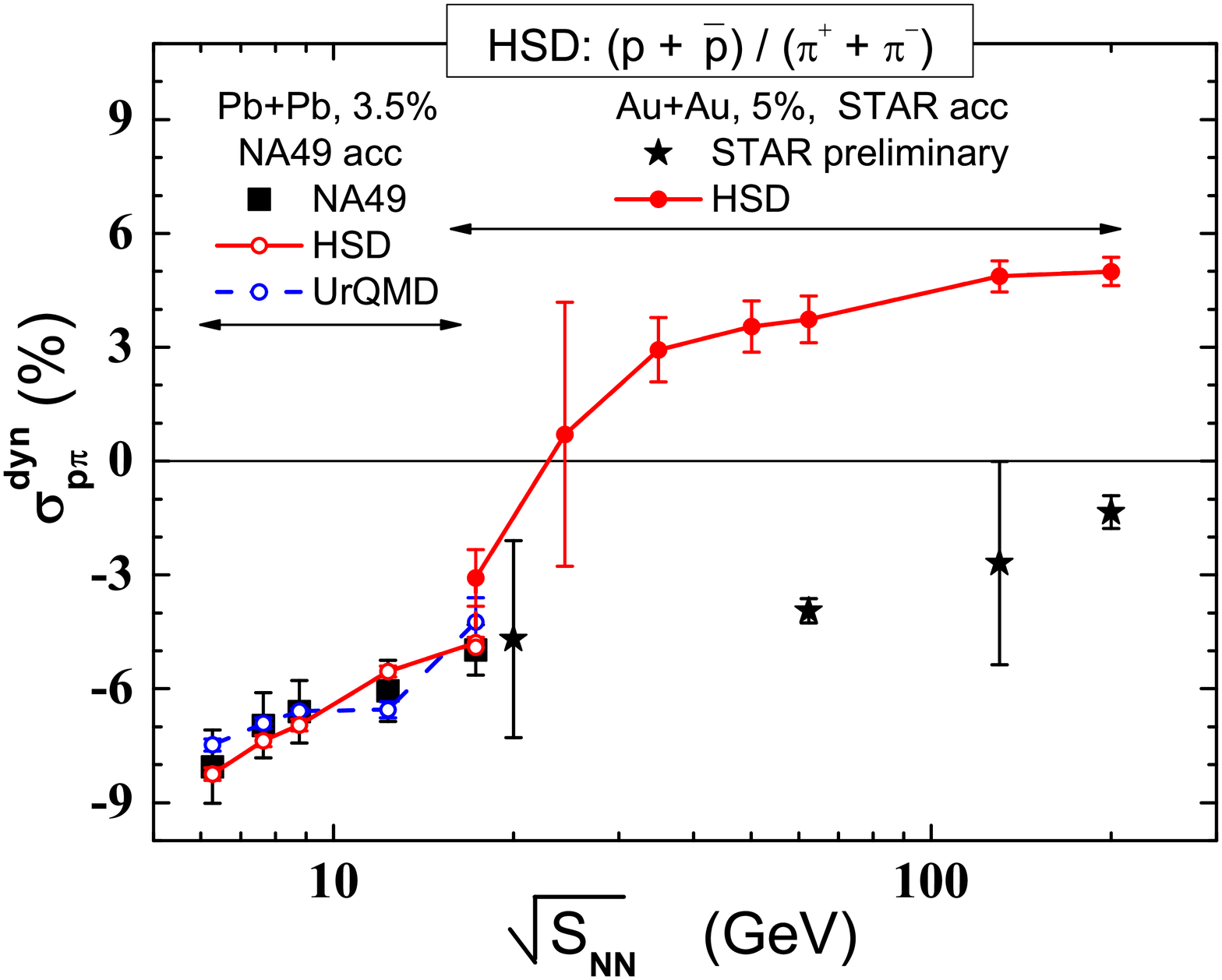,width=12cm}
\epsfig{file=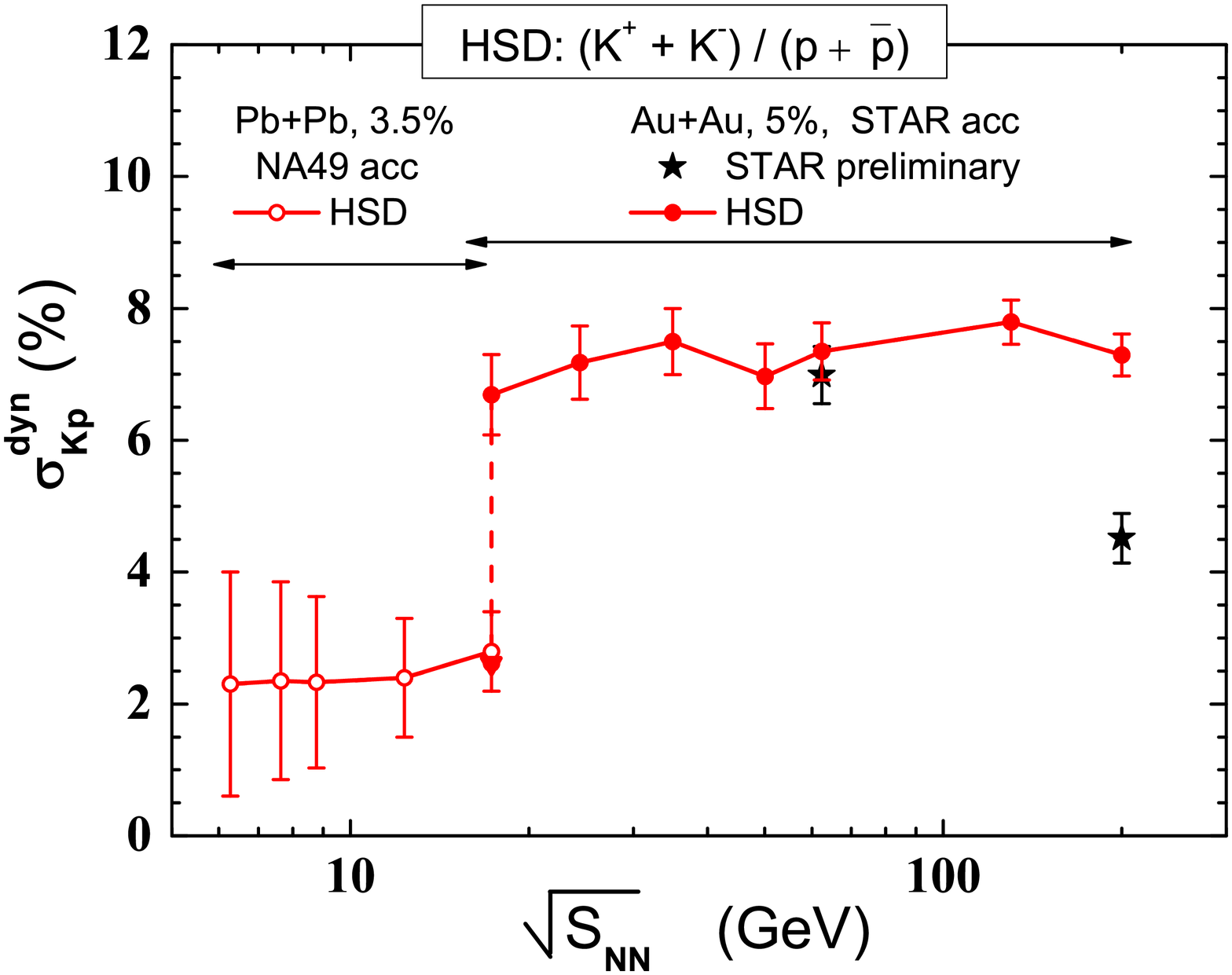,width=12cm}
\caption{(Color online)  
The solid lines present the HSD results of $\sigma^{dyn}$ (\protect\ref{sigmadyn}) 
for the $(p+\bar{p})/(\pi^++\pi^-)$ ({\it upper panel}) and $(K+K^-)/(p+\bar{p})$ ({\it lower panel}) ratios. 
The data are from Refs. \cite{NA49ratio,STAR-PPi,KP}.} 
\label{fig5}
\end{figure}

A comparison of the SM results for fluctuations in different
ensembles with the data looks problematic at present; 
the same is true for most other theoretical models. 
This is because of difficulties in implementing the experimental acceptance and centrality selection 
which, however, can be taken into account in the transport approach. 
In order to compare the HSD calculations with the measured data, the experimental cuts are
applied for the simulated set of HSD events.
In Fig. \ref{fig5} the HSD results of $\sigma^{dyn}$ (\ref{sigmadyn}) 
for the $p/\pi$ and $K/p$ ratios are shown in comparison with 
the experimental data by the NA49 Collaboration at the SPS~\cite{NA49ratio}
and the preliminary data of the STAR Collaboration at RHIC~\cite{STAR-PPi}. 
The results of the UrQMD calculations for $\sigma^{dyn}_{p\pi}$ 
at the SPS energies (from Ref.~\cite{NA49ratio}) are also shown by the dashed line.

For the SPS energies we use the NA49 acceptance tables from Ref.~\cite{NA49ratio}.
For the RHIC energies we use the following cuts: in pseudorapidity, $|\eta|<1$, 
and in transverse momentum, $0.2<p_T<0.6$~GeV/c for kaons and pions 
and $0.4<p_T<1$~GeV/c for protons~\cite{STAR,STAR-PPi,KP}. 
We note also, that HSD results presented in Fig.~\ref{fig5} correspond to the centrality
selection as in the experiment: the NA49 data correspond to the
3.5\% most central collisions selected via veto calorimeter,
whereas in the STAR experiment the 5\% most central events with
the highest multiplicities in the pseudorapidity range
$|\eta|<0.5$ have been selected.

At the SPS energies the HSD simulations lead to negative values of
$\sigma^{dyn}$ for the proton to pion ratio. 
This is in agreement with the NA49 data in Pb+Pb collisions. On the other hand HSD gives
large positive values of $\sigma^{dyn}_{p\pi}$ at RHIC energies
which strongly overestimate the preliminary STAR data for Au+Au collisions \cite{STAR-PPi}. 
For $\sigma^{dyn}_{Kp}$ only preliminary STAR data in Au+Au collisions are available~\cite{KP} 
which demonstrate a qualitative agreement with the HSD results (Fig.~\ref{fig5}). 
The HSD results for $\sigma^{dyn}_{Kp}$ show a weak energy dependence in both SPS and RHIC energy regions. 
A peculiar feature is, however, a strong `jump' between the SPS and RHIC values 
seen in the {\it lower panel} of Fig.~\ref{fig5},
in the HSD calculations which is caused by the different acceptances in the SPS and RHIC measurements.

The influence of the experimental acceptance is clearly seen at
160 A GeV where a switch from the NA49 to the STAR acceptance leads
to the jump in $\sigma^{dyn}_{Kp}$ by 3\% - lower panel of Fig.~\ref{fig5}. 
On the other hand, our calculations
for Pb+Pb (3.5\% central) and for Au+Au (5\% central) collisions
- performed within the NA49 acceptance for  both cases at 160 A GeV -
shows a very week sensitivity of $\sigma^{dyn}_{Kp}$ on the actual
choice of the collision system and centrality -- 
cf. the coincident open circle and triangle at 160 A GeV in the lower panel of Fig. \ref{fig5}.

\section{Summary and conclusions}

We have studied the event-by-event fluctuations of the number of
protons (and anti-protons), the proton-pion and proton-kaon
correlations, and the fluctuations of proton to pion and kaon to
proton ratios in central Pb+Pb and Au+Au collisions from low SPS
up to top RHIC energies. 
The analysis has been performed within the statistical
hadron-resonance gas model for different statistical ensembles --
the grand canonical ensemble (GCE), canonical ensemble (CE), and
micro-canonical ensemble (MCE) -- and in the
Hadron-String-Dynamics transport approach. We have found that the
HSD results at SPS energies are close to those in the CE and
MCE statistical model. 
This indicates a dominant role of
resonance decays and global conservation laws at low energy
nucleus-nucleus collisions. On the other hand, substantial
differences in HSD and statistical model results have been
observed at RHIC energies which can be attributed to non-equilibrium
dynamical effects in the HSD simulations.

The HSD results for $\sigma^{dyn}_{p\pi}$ appear to be close to
the NA49 data at the SPS. The data for $\sigma^{dyn}_{Kp}$ in
Pb+Pb collisions at the SPS energies will be available soon and allow for further insight. 
A comparison of the HSD results with preliminary STAR data in Au+Au
collisions at RHIC energies are not fully conclusive:
$\sigma^{dyn}$ from HSD calculations are approximately in
agreement with data \cite{KP} for kaon to proton ratio, but
overestimate the experimental results \cite{STAR-PPi} for proton
to pion ratio. New data on event-by-event fluctuations in Au+Au at
RHIC energies will help to clarify the situation.


\vspace{0.5cm} {\bf Acknowledgements}

We like to thank M.~Bleicher, W.~Cassing, M.~Ga\'zdzicki,
W.~Greiner, C.~H\"ohne, D.~Kresan, M.~Mitrovski, T.~Schuster,
R.~Stock, H.~Str\"obele, G.~Torrieri, G.~Westfall and S.~Wheaton for useful
discussions. This work was in part supported by the Program of
Fundamental Researches of the Department of Physics and Astronomy
of National Academy of Sciences, Ukraine.


\end{document}